# Evaluation of neutronic parameters for RITM-200 reactor unit considering $(^{238}U+^{235}U)O_2$, $(^{232}Th+^{235}U)O_2$ and $(^{232}Th+^{233}U)O_2$ dispersed fuel using MCU-PTR

Alhassan S., Beliavskii S.V., Nesterov V. N.,

National Research Tomsk Polytechnic University,

**Abstract.** Simulation tools have become an integral tool in the analysis of neutronic parameters of reactor units. These simulation tools are built to solve the neutron transport equation. In this article, the MCU-PTR simulation tool is used in the evaluation of the possibility of achieving extra-long fuel lifetime in the RITM-200 reactor unit with consideration to the fuel composition. The three dispersed fuel compositions $(^{238}U+^{235}U)O_2$, $(^{232}Th+^{235}U)O_2$ and $(^{232}Th+^{233}U)O_2$ are simulated at varying fuel element diameter of 4.9 mm to 8.9 mm with 1 mm interval. From the simulation, the dependence of reactivity margin on the fuel burnup of the three dispersed fuels were established. The dependence of the effective multiplication factor ($k_{eff}$) on the reactor operation time for the three dispersed fuel composition were also determined. From the results, it can be deduced that transition of the fuel from $(^{238}U + ^{235}U)O_2$ to $(^{232}Th + ^{235}U)O_2$ results in a 12% increase in the fuel burnup and no change in fuel campaign. While the transition from $(^{238}U + ^{235}U)O_2$ to $(^{232}Th + ^{233}U)O_2$ leads to a 32.4% increase in the fuel campaign and 45.6% increase in the fuel burnup. It has also been shown that the optimal fuel element diameter is 7.9 mm for $(^{232}Th + ^{233}U)O_2$ dispersed fuel. At this diameter, it is possible to increase the duration of the fuel campaign by 85.3% and achieve a fuel burnup of 51.9% in comparison to the design fuel diameter of 6.9 mm for $(^{238}U + ^{235}U)O_2$ dispersed fuel.

### 1.1 Introduction

The Monte-Carlo Universal software (MCU-PTR) developed by the Kurchatov Institute is a Monte-Carlo code with a combinatorial three-dimensional geometry that simulates radiation transport with continuous energy spectrum, calculation of criticality, spatial and energy distribution calculations of neutron fields, the ability to change nuclide composition during fuel lifetimes calculates etc [1, 2].

This Monte-Carlo software like the other types uses a library of evaluated data (based on BNAB-93) [3], also ACE/MCU (based on ENDF/B-VII.0) library could be used. The three-dimensional combinatory geometry of the software uses MCU/LATTICE and MCU/NET to place elements in nodes of regular lattice in order to model the RITM-200 reactor core. For every nuclide composition change, a $3\times10^6$ neutron histories were simulated [4 - 7]. MCU/BURN5 module allows for nuclide changes set with an integration time step of 50 effective full power days (EFPD) until the fuel lifetime end when the reactivity margin reaches 0.

Below is figure 1 illustrating the RITM-200 reactor core and fuel assembly layout as derived from the visual editor of the MCU-PTR software.

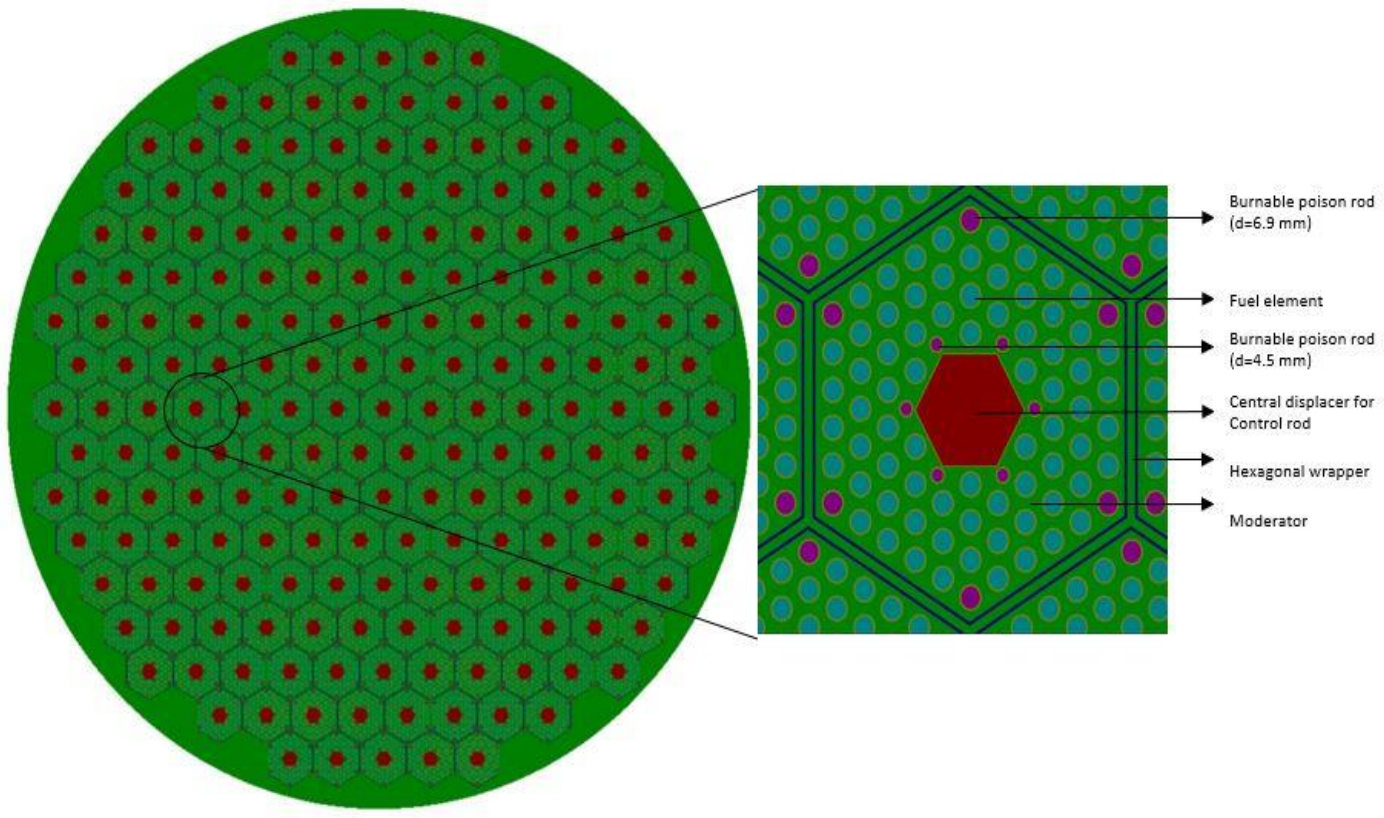

Figure 1 Horizontal layout of the RITM-200 reactor core and fuel assembly

### 1.2 MCU-PTR simulation of the RITM-200 reactor unit

In this simulation tool, the input data contains defined concentrations and volumes of all contents of the reactor core. The fissile nuclide enrichment for all fuel compositions is 19.2wt%. The dispersed fuel density used in the simulation is the same for all cases. These materials include, the wrapper, water (coolant), fuel content; $(^{238}U+^{235}U)O_2$, $(^{232}Th+^{235}U)O_2$ and $(^{232}Th+^{233}U)O_2$, the vacuum defined by the code as aluminium, burnable absorbers and the clad. The geometer of the reactor unit is designed with zones, bounded surfaces and planes. In RITM-200, the 199 fuel assemblies which house 72 fuel rods and12 burnable absorber rods in each fuel assembly are created. Each material is defined by its temperature, nuclide composition and concentration. The parameters of the reactor are set out for the simulation in the form of a heterogeneous core.

Fissile utilization is estimated as [8]:

$$B_{\text{fiss}} = \frac{B}{C_{\text{fiss}}}, \tag{1}$$

where $B$ denotes the fuel burnup, MW·d/kg$_{HM}$; $C_{fiss}$ denotes the fissile nuclide content.

The fissile nuclide content in percentage form is calculated as follows [9]:

$$C_{\text{fiss}} = \frac{\sum_{i}^{n} M_i \cdot N_i}{\sum_{k}^{m} M_k \cdot N_k}, \text{wt\%} \tag{2}$$

where $M_i$, $M_k$, connotes the molar mass of the $i$-th fissile material and the $k$-th heavy material, g/mol; $N_i$, $N_k$ refers to the nuclear concentration of the $i$-th fissile material ($^{233}U$, $^{235}U$, $^{239}Pu$, and $^{241}Pu$) and the $k$-th heavy material ($^{232}$Th, $^{233}$U, $^{234}$U, $^{235}$U, $^{238}$U, $^{239}$Pu, $^{240}$Pu, $^{241}$ Pu, and $^{242}$Pu); *n* identifies the number of fissile materials; *m* signifies the number of heavy materials.

The contribution of each *i*-th energy group of neutrons into fission ($NGC_i$) was also estimated using the equation below [10]:

$$\mathrm{NGC_i} = \frac{\Sigma_{f,i} \cdot \delta}{\sum_{i=1}^{26} \Sigma_{f,i} \cdot \delta_i}, \% \quad (3)$$

where $\Sigma_{f,i}$ signifies the macroscopic cross-section of fission in the *i*-th energy group, $cm^{-1}$; $\delta_i$ denotes the ratio of the neutron flux in the *i*-th group to the total neutron flux.

**2.1 Results of calculations of the duration of the nuclear fuel campaign of the heterogeneous core using the MCU-PTR program**

The results of the calculation of the effective multiplication against reactor operation time is observed in figure 2. Superior performance of ($^{232}$Th+$^{233}$U)$O_2$ dispersed fuel attains a neutron multiplication factor ($k_{eff}$) of 1.6330 over a reactor operation time of 2250 effective days while ($^{238}$U+$^{235}$U)$O_2$ dispersed fuel is the least performing fuel with neutron multiplication factor of ($k_{eff}$) 1.3671 over a reactor operation time of 1700 effective days.

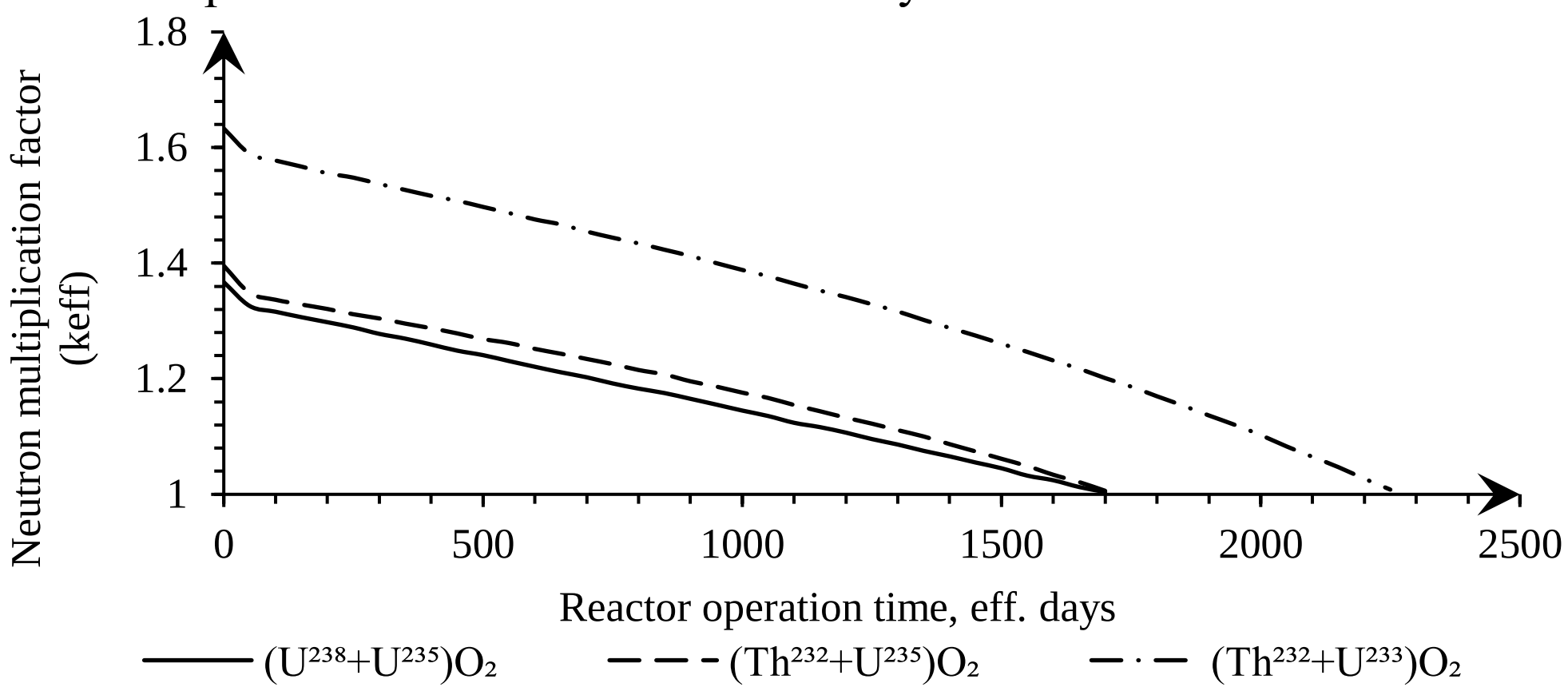

Figure 2 – Effective multiplication with the reactor operation time for different dispersed nuclear fuel

For the graph of the dependence of reactivity margin of the three dispersed fuel on fuel burnup as seen in figure 3 below, a similar trend is observed with the superior performance attained by ($^{232}$Th+$^{233}$U)$O_2$ fuel composition. The initial reactivity margins of ($^{238}$U+$^{235}$U)$O_2$ and ($^{232}$Th+$^{235}$U)$O_2$ are 0.2685 and 0.2834 respectively. ($^{232}$Th+$^{235}$U)$O_2$ achieves a higher fuel burnup of 108.1 MWd/ kgHM compared to 96.61 MWd/ kgHM by ($^{238}$U+$^{235}$U)$O_2$ dispersed fuel.

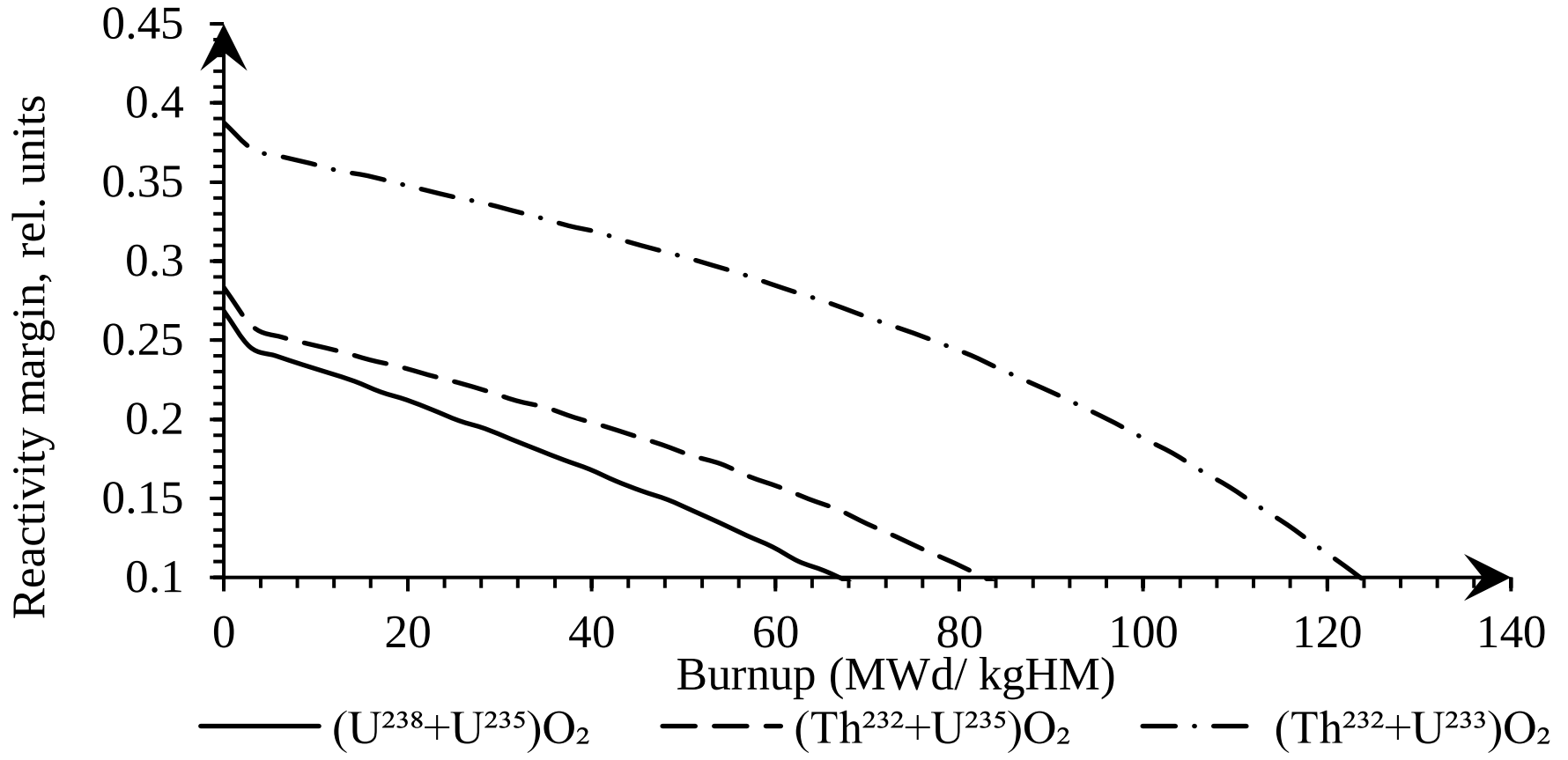

Figure 3 – The dependence of the reactivity margin on the fuel burnup

### 3.1 Calculation of fuel campaign of RITM-200 Reactor at different fuel diameters using MCU-PTR

The calculation of the fuel campaign at different fuel element diameters showed the possibility of optimizing the fuel element to achieve higher performance than the current design fuel element diameter of 6.9 mm. From the results of the MCU-PTR simulation, at 7.9 mm fuel element diameter the best fuel performance is achieved for the thorium-uranium fuel compositions. For $(^{238}U+^{235}U)O_2$ fuel composition, a continuous increment is observed in the fuel lifetime to the diameter of 6.9 mm at 1700 effective days of reactor operation. A marginal increase in fuel lifetime is observed from 6.9 mm to 8.9 mm at 2100 effective days. $(^{232}Th+^{235}U)O_2$ dispersed fuel attains a similar trend of increment from 600 to 2550 effective days over the fuel element diameter from 4.9 to 8.9 mm. A superior performance is observed from $(^{232}Th+^{233}U)O_2$ dispersed fuel with the fuel lifetime of 4150 effective days at 8.9 mm. This is shown in figure 4 below.

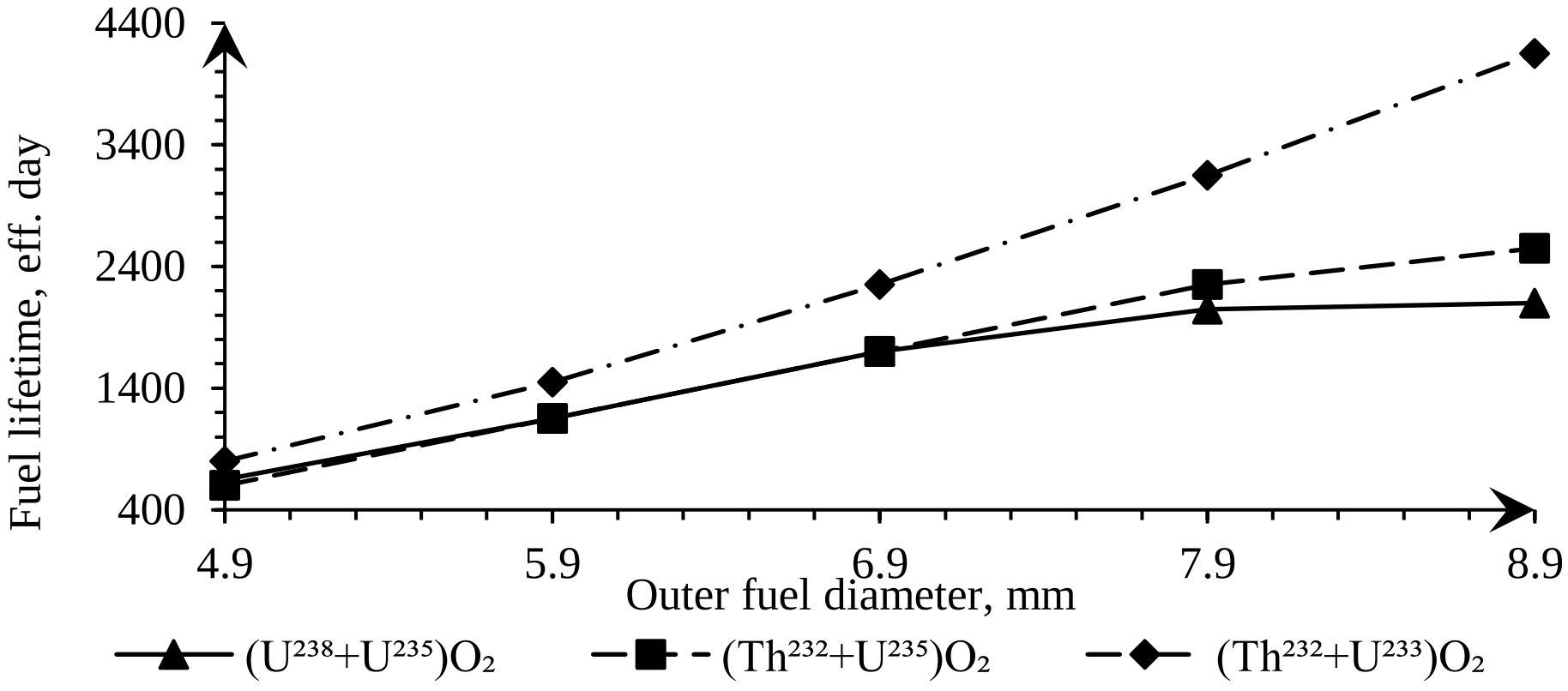

Figure 4 – Dependence of the duration of the nuclear fuel campaign on the diameter of fuel rods with different dispersion fuels

The optimal fuel element diameter is observed at 7.9 mm based on the results of the reactivity margin against outer fuel diameter of the three dispersed fuel compositions. From the results of the MCU simulation, the highest reactivity margin for both $(^{238}U+^{235}U)O_2$ and $(^{232}Th+^{235}U)O_2$ is reached at 6.9 mm. In the $(^{232}Th+^{233}U)O_2$ fuel composition, a constant reactivity margin is seen between the fuel element diameters 6.9 and 7.9 mm. Hence, since the graph from figure 4 indicates increasing outer fuel diameter leads to increasing fuel lifetime, the best performance of the fuel composition will be attained at 7.9 mm fuel element diameter. The graph of the dependence of the reactivity margin on the diameter of fuel rods with different dispersed fuels is shown in figure 5 below.

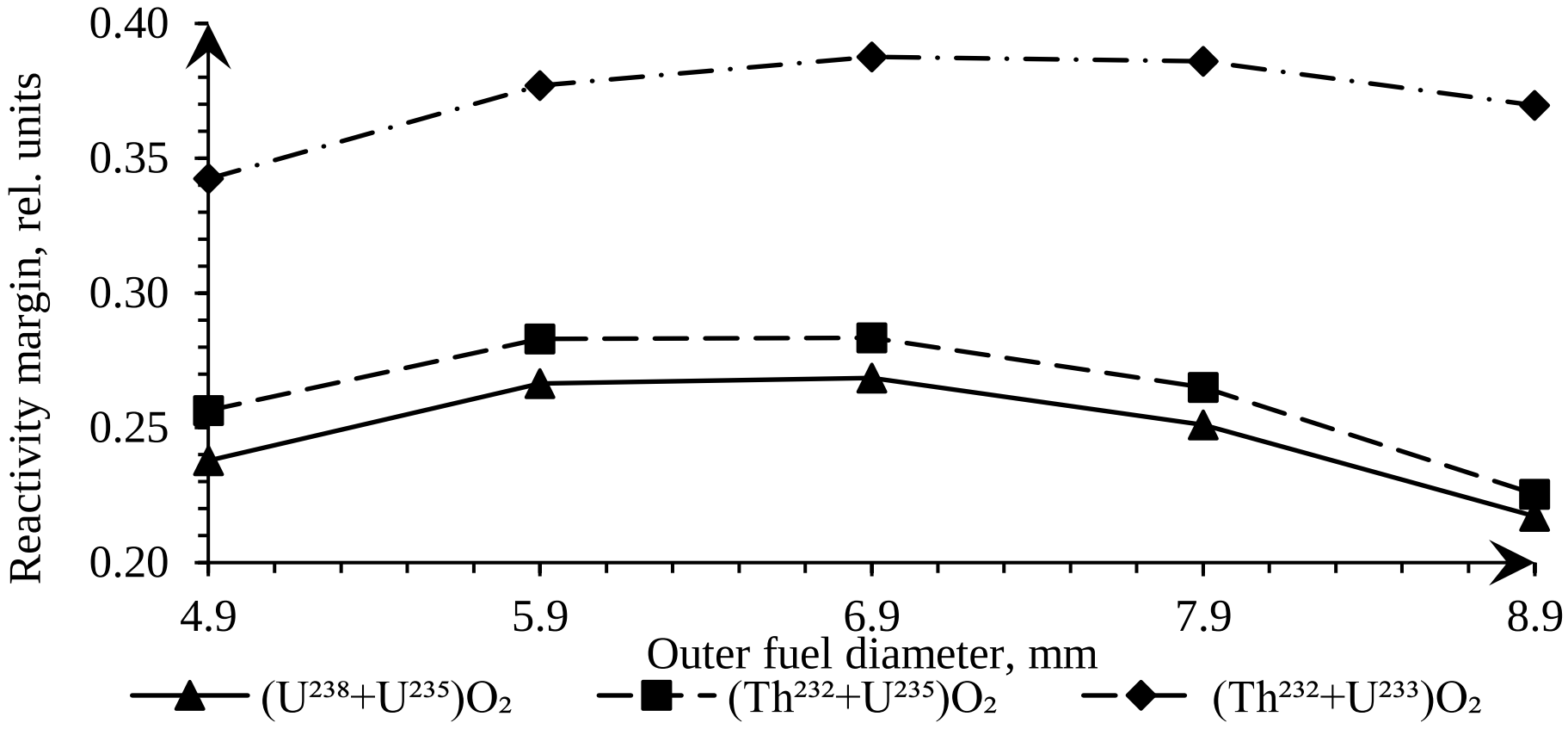

Figure 5 – The dependence of the reactivity margin on the diameter of fuel rods with different dispersion fuels

## Conclusion

From the MCU simulation of the neutronic parameters of the RITM-200, it can be deduced that transition of the fuel from $(^{238}U + ^{235}U)O_2$ to $(^{232}Th + ^{235}U)O_2$ results in a 12% increase in the fuel burnup and no change in fuel campaign. Similarly, the transition from $(^{238}U + ^{235}U)O_2$ to $(^{232}Th + ^{233}U)O_2$ leads to a 32.4% increase in the fuel campaign and 45.6% increase in the fuel burnup. With an optimal value of the fuel element diameter at 7.9 mm, it is possible to increase the duration of the fuel campaign by 85.3% and achieve a fuel burnup of 51.9% for $(^{232}Th + ^{233}U)O_2$ dispersed fuel in comparison to the design fuel diameter of 6.9 mm for $(^{238}U + ^{235}U)O_2$ dispersed fuel.